\documentclass[letterpaper]{article} 
\usepackage{aaai24}  
\usepackage{times}  
\usepackage{helvet}  
\usepackage{courier}  
\usepackage[hyphens]{url}  
\usepackage{graphicx} 
\usepackage{natbib}  
\usepackage{caption} 
\usepackage{algorithm}
\usepackage{algorithmic}

\usepackage{newfloat}
\usepackage{listings}
\DeclareCaptionStyle{ruled}{labelfont=normalfont,labelsep=colon,strut=off} 
\floatstyle{ruled}
\newfloat{listing}{tb}{lst}{}
\floatname{listing}{Listing}
\usepackage{xcolor,colortbl}

\usepackage{xcolor}

\usepackage[en-US]{datetime2}

\DTMsavetimestamp{creation}{2024-04-01T00:00:00-05:00}
\date{\DTMusedate{creation}}

\usepackage{subcaption}
\usepackage{tabularx}
\usepackage{booktabs}
\usepackage{comment}
\usepackage{multirow}
\usepackage{amsmath}
\usepackage{soul}
\usepackage{xspace}
\usepackage{xcolor,colortbl}
\usepackage[main=english,russian,czech]{babel}

\definecolor{Gray}{gray}{0.85}
\definecolor{LightCyan}{rgb}{0.88,1,1}

\newcolumntype{a}{>{\columncolor{Gray}}r}
\usepackage{xcolor}

\newcommand*{\escape}[1]{\texttt{\textbackslash#1}}
\newcommand*{\escapeI}[1]{\texttt{\expandafter\string\csname #1\endcsname}}

\usepackage{csquotes}

\title{Partial Mobilization: Tracking Multilingual Information Flows Amongst Russian Media Outlets and Telegram}
\author {
    Hans W. A. Hanley and 
    Zakir Durumeric\\
}
\affiliations {
    Stanford University \\
    hhanley@cs.stanford.edu, zakir@cs.stanford.edu
}

\begin{document}

\title{Partial Mobilization: Tracking Multilingual Information Flows amongst Russian Media Outlets and Telegram}

\maketitle
\begin{abstract}
In response to disinformation and propaganda from Russian online media following the invasion of Ukraine, Russian media outlets such as Russia Today and Sputnik News were banned throughout Europe. To maintain viewership, many of these Russian outlets began to heavily promote their content on messaging services like Telegram. In this work, we study how 16~Russian media outlets interacted with and utilized 732~Telegram channels throughout 2022. Leveraging the foundational model MPNet, DP-Means clustering, and Hawkes processes, we trace how narratives spread between news sites and Telegram channels. We show that news outlets not only propagate existing narratives through Telegram but that they source material from the messaging platform. For example, across the websites in our study, between 2.3\% (ura.news) and 26.7\% (ukraina.ru) of articles discussed content that originated/resulted from activity on Telegram. Finally, tracking the spread of individual topics, we measure the rate at which news outlets and Telegram channels disseminate content within the Russian media ecosystem, finding that websites like ura.news and Telegram channels such as @genshab are the most effective at disseminating their content. 
\end{abstract}

\section{Introduction}

On February 24, 2022, the Russian Federation invaded
Ukraine. In the buildup to and following the initial invasion, Russian state media conducted massive information campaigns justifying the Russian state's invasion as a  ``special military operation'' to ``liberate'' the people of Ukraine~\cite{Wong2022}. In response, the EU and the UK among others banned or otherwise censored Russian news media. To circumvent this censorship, Russian outlets like RT and Sputnik News began to redirect users to and promote their content on the messaging app Telegram; 
ultimately causing Telegram to become one of the main platforms for Russian propaganda~\cite{Bergengruen2022}. When Telegram eventually succumbed to pressure to de-platform prominent Russian outlets (\textit{e.g.}, rtnews), several Russian state media outlets created new Telegram channels (\textit{e.g.}, swentr [rtnews backward])\footnote{\url{https://web.archive.org/web/20220304200814/https://www.rt.com/russia/551256-how-access-rt-censorship-bypass/}} and found other ways to circumvent the bans. However, while there has been extensive reporting on the Russian state media's usage of Telegram~\cite{Bergengruen2022}, there has been no systematic study of how information flows between Russian media and Telegram. 

In this paper, we document the increased usage of Telegram by Russian outlets and present the first programmatic and multilingual study of the spread of news content amongst and between Russian news sites and Telegram. To do this, we crawl and gather content published between January 1 and September 25, 2022, from 16~Russia-based news sites (215K articles) and the 732~Telegram channels hyperlinked by these Russian news sites (2.48M~Telegram messages). Leveraging a multilingual version of the large foundational model MPNet~\cite{song2020mpnet}, fine-tuned on semantic search, we perform semantic similarity analyses of the content spread between and amongst these news platforms and our corresponding set of English-language, Russian-language, and Ukrainian-language Telegram channels. Further, by improving upon an online and parallelizable non-parametric version of the K-Means algorithm, we cluster our dataset into fine-grained \textit{topic clusters} to understand the topics discussed. 



We find that much of the content shared between Telegram and Russian websites concerns the war in Ukraine and Western sanctions on Russia. By performing this same clustering on semantic content specific to only Russian news websites or only Telegram channels, we find an emphasis on the day-to-day machinations (\textit{e.g.}, bombings of particular bridges) of the Russo-Ukrainian War on Telegram that contrasts with a focus on US politics specific on Russian news outlets. Next, we track the spread of topics amongst and between Telegram and Russian news media. We find that 33.2\% of distinct topics discussed on our set of Telegram channels, making up 24.3\% of all Telegram messages in our dataset, were first published in Russian news articles. In contrast, 13.9\% of topics on Russian news websites, comprising 18.4\% of all the text (all messages from Telegram and paragraphs from Russian state media) on our set of Russian websites, were posted first on Telegram. Telegram-originated messages comprise a particularly large amount of content on news websites like {waronfakes.com} (28.2\% of text), {ukraina.ru} (27.9\%), and {ura.news} (25.6\%) that all maintain large Telegram presences. Applying time-series analysis with the Hawkes processes on our topic clusters, we then estimate the percentage of content on each platform that was influenced by activity on other platforms. While some websites like ura.news had relatively low amounts of content (2.3\%) flowing from Telegram, others such as ukraina.ru had much larger (26.7\%) percentages of their content possibly influenced by activity on Telegram. Finally, we analyze how quickly topics spread amongst our set of news websites and onto Telegram channels, finding that websites like ura.news, ukraina.ru, and news-front.info, and the Telegram channel @genshab's topics flow most quickly to other platforms.


Our work presents one of the first in-depth analyses of semantic content and semantic similarity among and between Russian state media websites and Telegram channels. We show that by leveraging multilingual models, we can programmatically track the spread of topics and ideas across platforms. As misinformation and propaganda increasingly spread on messaging platforms like Telegram and WhatsApp, we hope that our work can serve as the basis for future studies about the spread of misinformation online.
\section{Background and Related Work\label{sec:background}}

\textbf{\textit{Telegram.}} Telegram, started in 2013, is a messaging platform ~\cite{baumgartner2020pushshift} with 700 million monthly users~\cite{Singh2022} as of June 2022. Similar to WhatsApp and Facebook Messenger, users on Telegram can share messages, images, and videos. However, in addition to private conversations, Telegram users can create one-to-many public channels (where only channel creators and administrators can post content) to which other users can subscribe. Within this work, we focus on these administrator-run channels. 

With a free-speech ethos, Telegram is a platform where extremist content, misinformation, and propaganda can thrive~\cite{walther2021us}. Both~\citet{hohn2022belelect} and~\citet{baumgartner2020pushshift}, for instance, develop public datasets for specifically analyzing the spread of misinformation on Telegram.~\citet{urman2022they} similarly examine far-right Telegram networks. Following the ban of many Russian news outlets throughout Europe, several Russian outlets have as a result turned to Telegram to spread their content, with Russia Today and Sputnik News even having pages dedicated to showing users how to download Telegram~\cite{bovet2022organization, Bergengruen2022}.  

\vspace{2pt}
\noindent
\textbf{\textit{Types of Unreliable Information.}}
Unreliable information can take the form of \textit{misinformation}, \textit{disinformation}, and \textit{propaganda}, among other types~\cite{jack2017lexicon}. \textit{Misinformation} is any information that is false or inaccurate regardless of the intention of the author. \textit{Disinformation}, in contrast, is false and inaccurate information spread with the express and deliberate purpose to mislead~\cite{jack2017lexicon}. Similar to disinformation, \textit{propaganda} refers to ``deliberate, systematic information campaigns, usually conducted through mass media forms'' regardless of whether the information is true or false ~\cite{jack2017lexicon}. Within this work, we do not make distinctions between different types of information flowing from Russian propaganda outlets, instead focusing on their overall use of Telegram and their relationships with one another. 

\vspace{2pt}
\noindent
\textbf{\textit{Russian Propaganda.}}
While we do not examine specific Russian propaganda stories, instead focusing on the different Russian platforms' use of Telegram, several other works have shown the widespread influence of Russian propaganda. For instance, following the 2016 US election, Badawy et~al.\ found that Russian bots propagated US pro-conservative and divisive messages,  especially towards users in the US South~\cite{badawy2018analyzing}. A similar study from the RAND corporation identified two communities of over 40,000 users on Twitter that promoted anti- and pro-Russian stories throughout Eastern Europe~\cite{helmus2018russian}. Finally, Hanley et~al.~\cite{hanley2022happenstance,hanley2022special} examined the spread of Russian propaganda to US and Chinese social media.

\vspace{2pt}
\noindent
\textbf{\textit{Multilingual Analyses of News and Misinformation.}}
As machine learning and natural language processing tools have improved, various works have taken multi-modal and multilingual approaches to detect and track news, disinformation, and propaganda. Using a multilingual BERT model,~\citet{panda-levitan-2021-detecting} analyze the spread of COVID-19 misinformation in Bulgarian, Arabic, and English. We note that multilingual models have been shown to suffer from reduced performance, especially for rarer languages~\cite{joshi2020state}. However, as shown by~\citet{verma2022overcoming}, multimodal approaches and investment in the training of more robust models can help ameliorate many of these issues.



\begin{table}
\centering
\fontsize{9pt}{6.5pt}
\selectfont
\setlength{\tabcolsep}{1pt}
\begin{subtable}[t]{0.23\textwidth}
\begin{tabularx}{\columnwidth}{Xrr}
\textbf{Platform} &\textbf{Art.} & \textbf{Emb. } \\\midrule
Telegram & -- &  2.48M \\
geopolitca.ru &544 & 4.22K\\
global-research.ca &5.90K & 158K \\
govorit-moskva.ru & 57.9K & 224K \\
journal-neo.org & 2.51K & 15.4K\\
katehon.com & 4.39K & 45.8K \\
lug-info.com & 1.22K & 8.92K \\
news-front.info & 29.0K & 262K\\

 &  &  \\
\bottomrule
\end{tabularx}
\hfill
\end{subtable}
\begin{subtable}[t]{0.23\textwidth}
\begin{tabularx}{\columnwidth}{Xrr}
\textbf{Platform} &\textbf{Art.} & \textbf{Emb.} \\\midrule
rbc.ru &16.0K & 179K \\
rt.com &12.7K & 138K\\
southfront.org & 7.79K& 79.0K \\
sputnik-news.com &18.2K & 183K \\
strategic-culture.org & 2.17K  & 25.1K\\
tass.com  & 9.11K & 44.9K\\
ukraina.ru & 17.4K & 193.6K \\
ura.news & 7.92K & 125.6K\\
waronfakes.com & 2.48K & 10.9K \\ 
\bottomrule
\end{tabularx}
\end{subtable}
\caption{\label{tab:dataset} The number of articles and embeddings extracted from Telegram and our set of 16~Russian media websites. Articles are embedded on a paragraph level while each Telegram individual message is embedded.} 
\end{table}

\section{Datasets\label{sec:datasets}}
We use several datasets to understand the interaction amongst and between Russian news media sites and Telegram. We provide an overview of these datasets here.

\vspace{2pt}
\noindent
\textbf{\textit{Russian Propaganda and State Media\label{sec:data-eng}.}}
Our study examines 11~Russian propaganda and state media websites previously analyzed by the US State Department and prior work~\cite{RussiaPillar2020, hanley2022happenstance,hanley2022special} (Table~\ref{tab:dataset}). In addition, we extend our set of websites to include five additional Russian-language news websites that have been documented to spread Russian propaganda~\cite{park2022voynaslov,Aleksejeva2022,von2017annual}: {lug-info.com}, {govoritmoskva.ru}, {ura.news}, {rbc.ru},  and {ukraina.ru}.

For each website, between August~20 and September~25 2022, utilizing the Go library {colly} and the Python library {Selenium}, we collect the articles published between January~1 and September~25, 2022. Specifically, for each website, we scrape 10~hops from the root page, collecting the article content published on each page (\textit{i.e.}, we collect all URLs linked from the homepage [1st hop], then all URLs linked from those pages [2nd hop], and so forth). To extract article content and associated metadata, we use the Python libraries {newspaper3k} and {htmldate}. For our Russian-language websites, {newspaper3k} was largely unable to collect article content. As a result, for each of these websites, we built a custom Python script to parse out article text from the scraped HTML\@. Altogether, we collect the content for 215,359~articles across the 16~websites (Table~\ref{tab:dataset}).

\vspace{2pt}
\noindent
\textbf{\textit{Telegram Channels.}}
To understand Russian platforms' use of Telegram, we curated the set of Telegram channels hyperlinked by our set of Russian news media websites throughout 2022. Specifically, from the pages of our Russian websites, we identified 802~Telegram channels. Removing private Telegram channels and those that were censored by Telegram or deleted (\textit{i.e.}, @rtnews), we were left with a total 732~Telegram channels. Then, as in prior work~\cite{hoseini2021globalization}, we scraped the public content of these channels from {www.t.me/channel\_name/s/} URLs. For each channel, we scrape all messages that were published between January~1 and September~25, 2022. Across the 732~channels, we collect a total of 2,592,772~Telegram messages (Table~\ref{tab:dataset}).

\section{Methodology\label{sec:methoddology}}
Having described our dataset, in this section, we outline several key algorithms that we utilize to track information flows across websites. Specifically, after detailing how we preprocess our data, we then describe how we determine the similarity of content between different websites/telegram channels after embedding \emph{text paragraphs} using a multilingual and fine-tuned version of the  MPNet~\cite{song2020mpnet} large foundation model. We finally specify the fine-grained clustering mechanism that we use to isolate and track specific topics across our set of websites.


\subsection{Preprocessing\label{sec:preprocessing}}
For each news article and Telegram message, before performing any analysis, we first remove all URLs, emojis, and HTML tags. We further discard any Telegram message that does not contain text or that has fewer than four words~\cite{hanley2022happenstance}. Altogether, we removed 115,208~messages that did not meet our criteria, leaving us with 2.48M~Telegram messages for our study (Table~\ref{tab:dataset}). For our news articles, we segment each article into its constituent paragraphs based on where newline ({\escape{n}}) or tab ({\escape{t}}) characters appear within the text. We note that this segmenting approach was successful for every website except {sputniknews.com} and {govoritmoskva.ru}. For both of these websites, we built custom scripts to segment their articles based on the text block elements specified in their HTML pages.

\subsection{Embedding Articles and Messages\label{sec:preprocessing}}

For every article and message, we embed its constituent paragraphs using our multilingual MPNet model. The specific model version we use is fine-tuned to the \textit{semantic search task} and trained to handle over 50~languages including English, Russian, and Ukrainian (the primary languages within our dataset).\footnote{\url{https://huggingface.co/sentence-transformers/paraphrase-multilingual-mpnet-base-v2}} This version of MPNet was trained so that texts with similar semantic content have higher cosine similarity.  We utilize the intuition, as in~\citet{hanley2022happenstance}, that when embedding text and performing subsequent analyses such as topic analysis, each embedding vector represents only one topic or idea. Any given article can contain multiple topics or ideas; thus we embed each paragraph of each article. Embedding paragraphs rather than sentences as in~\citet{hanley2022happenstance} enables us to obtain additional context while also obtaining an embedding for the (often) one topic/idea present within the paragraph. Altogether we embed 1.62M~paragraphs and 2.48M Telegram messages (57~minutes utilizing an Nvidia RTX A6000 GPU).  

Many of our articles and messages are not in English. Using the Python library {langdetect}, we find that 73.2\% of article paragraphs are in Russian, 22.3\% in  English, and 2.1\% in  Ukrainian. The remaining belong to an assortment of other languages. Similarly, 81.3\% of our Telegram dataset is in Russian, 6.0\% in English, and 5.3\% in Ukrainian. 



\vspace{-5pt}
\subsection{Comparing Semantic Content\label{sec:compare-semantic}}

We compare the semantic content of our embedded messages and paragraphs utilizing cosine similarity~\cite{song2020mpnet}. As found in previous works~\cite{hanley2022happenstance,song2020mpnet,bernard2022tracking}, a cosine similarity threshold between 0.60 and 0.80 can be utilized to determine whether two pieces of text are about the same topic. For instance, with the same model,~\citet{phan2022vietnamese}, found that a threshold near 0.715 achieved the best results. However, in order to further verify these past results, we perform our own evaluation to determine a threshold at which two messages/paragraphs can reliably be said to be about the same topic. 

We take two approaches to determine the appropriate threshold for considering two paragraphs to be about the same topic. First, we take 250 random paragraph pairs with similarities at various thresholds (\textit{i.e.}, for 0.60, messages/paragraphs with similarities between 0.59 and 0.61) and have an expert determine whether they are about the same topic as outlined in~\citet{hanley2022happenstance,soper2021semantic}. We perform this evaluation in monolingual settings for English and Russian (the two most prominent languages in our dataset) and a multilingual setting with English and Russian. Given that our expert could not speak Russian, we utilized Google Translate to translate our set of Russian paragraphs into English. As seen in Table~\ref{tab:similar-topic-precision}, our selected pre-trained model achieves a near 85.2\% topic-similarity precision at a threshold of 0.7 only for English. For Russian, and in a multilingual setting of English and Russian, our model only achieves this precision at a threshold of 0.8. This is in line with prior work~\cite{verma2022overcoming} that has found that multilingual models often over-perform in English while underperforming in rarer languages.

Second, to confirm our manual evaluation, we utilize~\citet{chen-etal-2022-semeval}'s dataset of 10K~multilingual article text paragraph pairs across 18~languages (including Russian and English) annotated for whether they were about the same news story. Analyzed across 7~criteria dimensions, this dataset labeled each paragraph pair on a scale of 1 (not about the same news story) to 4 (about the same news story).  Embedding each pair with our multilingual model, we find that 93.6\% of paragraph pairs with similarity above that at the threshold of 0.80 had very high similarity (a labeled score of 3.0 or higher). Similarly, 83.6\% of pairs below this threshold had lower similarity (lower than 3.0). Benchmarking our approach with this dataset, we confirm that the cosine similarity threshold of 0.80 is a strong indication that two paragraphs are about the same story. Throughout the rest of this work, when comparing two embeddings, we utilize a threshold of 0.8. This new evaluation matches previous works' evaluation of this particular model~\cite{huertas2021countering,phan2022vietnamese}. When two messages/paragraphs reach this threshold, we consider them to be \textit{similar} or to \textit{correspond} with one another.


\begin{table}
\centering
\fontsize{9pt}{8.5pt}
\selectfont

\begin{tabular}{r|rrr}
\textbf{Threshold} &\textbf{English } & \textbf{Russian } & \textbf{ English \& Russian} \\\midrule
{0.6} & 	48.0\%& 24.4\%& 28.8\%\\  
{0.7} & 	85.2\%  & 49.2\%&        62.4\%   \\  
{0.8} &  99.2\%    & 89.2\%&        89.6\% \\
{0.9} &$\sim$100.0\%     & $\sim$100.0\%&      $\sim$100.0\%  \\
\bottomrule

\end{tabular}
\caption{\label{tab:similar-topic-precision} Precision evaluation of whether embedded paragraphs/messages have the same topic at various thresholds and across different languages.} 
\end{table}

\vspace{2pt}
\noindent
\textbf{\textit{Computing Similarity Scores Between Platforms.\label{sec:similiarty-scores}}} Utilizing our threshold of 0.8, we compute the percentage of different platforms' messages/paragraphs that convey the same topic. This is such that we calculate the percentage of messages/paragraphs from one website whose topic/idea appears on another website (\textit{i.e.}, what percentage of {rt.com's} paragraphs also appear on {sputniknews.com} and conversely what percentage of {sputniknews.com's} paragraphs appear on {rt.com}). To consolidate the two similarity values into one average to approximate platform similarity, we take the geometric average of two returned percentages (We take the geometric average rather than arithmetic as the two numbers are largely non-independent~\cite{huntington1927sets}). 

\vspace{-5pt}
\subsection{Isolating Individual Topics\label{sec:cluster}}
In addition to identifying the similarity between individual messages/paragraphs and amongst websites, we further seek to identify individual topics/units of information within the Russian news ecosystem and track their spread. As in~\citet{hanley2022happenstance}, we utilize clustering to identify \textit{topics} present within our dataset. However, finding that the density-clustering approach utilized by~\citet{hanley2022happenstance} could not scale to our dataset, we utilize DPMeans to identify topic clusters.

DP-Means~\cite{dinari2022revisiting} is a non-parametric extension of the K-means algorithm that does not require the specification of the number of clusters \textit{a priori}. Within DP-Means, when a given datapoint is a chosen parameter $\lambda$ away from the closest cluster, a new cluster is formed, and that datapoint is assigned to it. As such, this enables us to specify how similar individual items must be to one another to be part of the same cluster. Similarly, because DP-Means is non-parametric, it does not require the specification of \textit{a priori} how many topics are present. 


We note that we make three key changes to the released version of parallelizable DP-Means:\footnote{\url{https://github.com/BGU-CS-VIL/pdc-dp-means}} (1) We cluster embeddings based on their cosine similarity with one another rather their Euclidean distances (also altering the cost function to consider cosine similarities); (2) We set new clusters to be formed whenever the message/paragraph is less than $\lambda =0.8$ similar to its nearest cluster; (3) We remove the random reinitialization of clusters in the released algorithm~\cite{dinari2022revisiting}; we find that this step often led to over-clustering given that many website paragraphs are slight variations of each other. These changes enable us to form highly fine-grained and semantically specific topic clusters both in monolingual and multilingual settings.


\vspace{-5pt}

\begin{figure*}
\begin{subfigure}{.32\textwidth}
  \centering
  \includegraphics[width=1\linewidth]{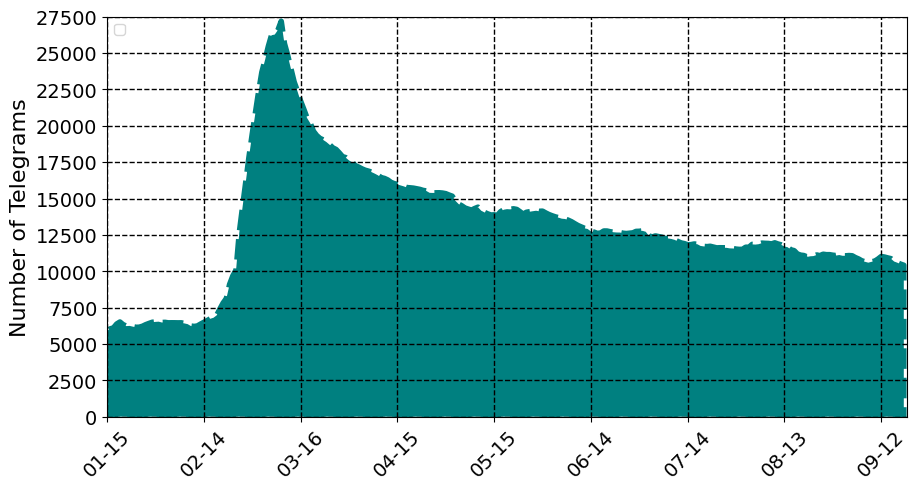}
  \caption{Number of Telegrams}
\label{fig:number-telegrams}
\end{subfigure}%
\begin{subfigure}{.32\textwidth}
  \centering
  \includegraphics[width=1\linewidth]{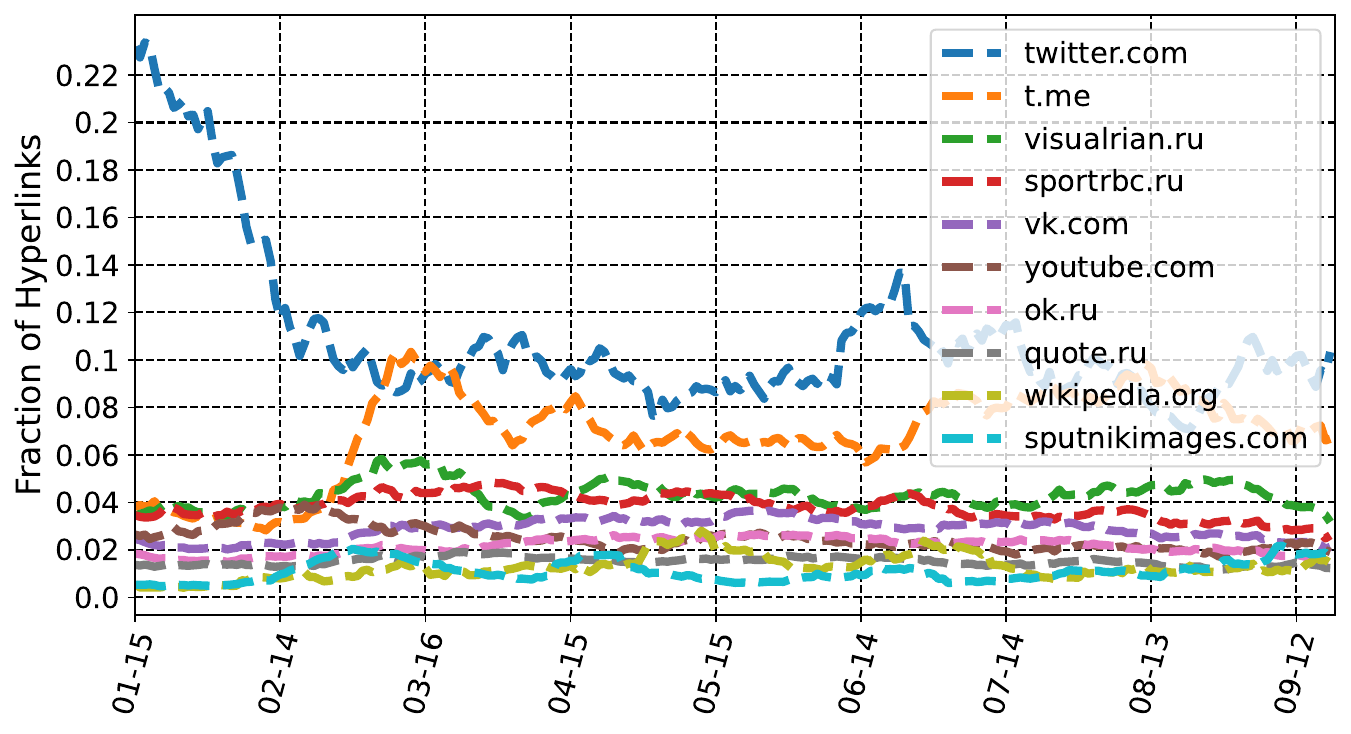}
  \caption{Hyperlinks from Websites}
  \label{fig:hyperlinks-websites}
\end{subfigure}%
\label{fig:website-domains-over-time}
\begin{subfigure}{.32\textwidth}
  \centering
  \includegraphics[width=1\linewidth]{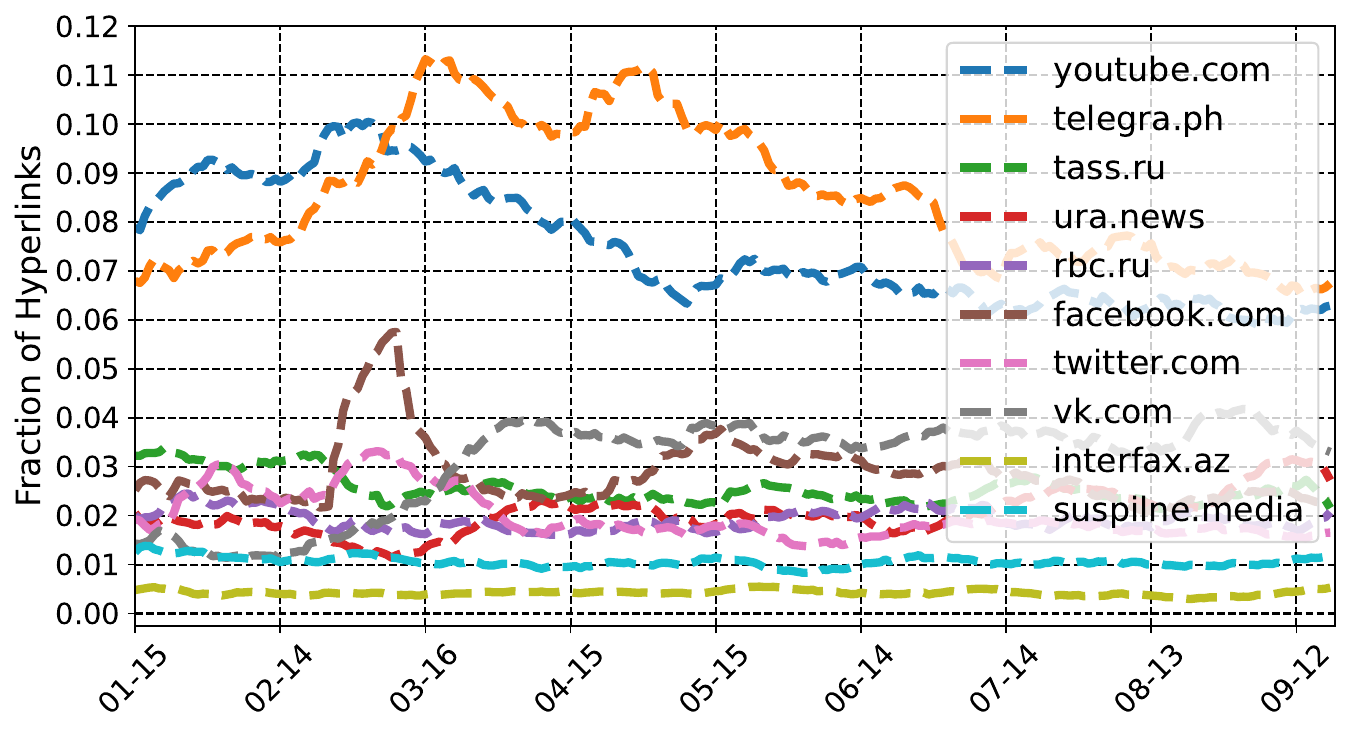}
  \caption{Hyperlinks from Telegram Channels}
  \label{fig:hyperlinks-telegrams}
\end{subfigure}%
\label{fig:activity}
\caption{Following, the Russian invasion of Ukraine, the number of Telegram posts in our set of 732 channels spiked from an average of 6,500 to over 27,500. Similarly, our set of Russian websites began to utilize Western media sites like Twitter less often while increasing their use of Telegram (Telegram links increased from 4\% of these websites' external hyperlinks to nearly 10\%). Our set of Telegram channels themselves began to steadily use platforms like YouTube less often.
}
\label{fig:narratives-from-telegram}
\vspace{-10pt}
\end{figure*}

\subsection{Estimating Platform Influence}
We utilize Hawkes processes to estimate the influence of the content on one platform on the content published on another. Hawkes processes are statistical models of event frequencies that account for the effect of other processes~\cite{linderman2015scalable}. This is such that a Hawkes process of one set of event frequencies can model the influence of another set of frequencies of its own frequency (\textit{e.g.}, rt.com mentioning a story may affect when and how much sputniknews.com reports on that same story). In this work, we fit the time series frequencies of particular events utilizing Gibbs sampling with settings as specified in past works~\cite{linderman2015scalable}.

Upon fitting our Hawkes processes, we note that the following weights are returned: (1) background rates from each process (also captures the influence of other processes not modeled [\textit{i.e.}, websites not included in our dataset]), (2) influence weights for each process to each other, and finally (3) shapes of the impulses of one process on another process \emph{and} itself. The background rate returned models the expected rate at which events will occur without influence from past events or from other processes (\textit{i.e.}, the rate at which a given platform writes about a given story or topic without influences from other websites or its previous set of articles on the given topic). The influence weights model how one event on one process (\textit{i.e.}, one mention of a given topic) influences the frequency of events on another process. For example, a weight of two from rt.com to sputniknews.com for a given topic means that each mention of that topic results in an \emph{expected} two more mentions by sputniknews.com. Finally, the shape of impulses from one process to another process models the probability that the \emph{expected} events caused by the first process will occur at a given point of time on the second process (\textit{i.e.}, one day after, two days after, \textit{etc...})

Utilizing these values and the steps laid out in past work~\cite{zannettou2018origins}, we calculate the influence of one platform on another (the percentage of a platform's messages or paragraphs that may have been caused by another platform) as well as the efficiency of different platforms in getting their content posted elsewhere (the amount of influence each platform has on another relative to the number of messages it itself posted).

\vspace{2pt}
\noindent
\textbf{Ethical Considerations}
We utilize only public data and follow ethical guidelines as outlined by others for scraping data~\cite{hanley2022no}.  We recognize that the Russo-Ukrainian War is an ongoing conflict and that sensitivity is paramount. We hope that our work provides objective insight into the behavior of Russian media outlets throughout the conflict.
\section{Behavior of Russian State Media and Associated Telegram Channels\label{sec:comparing} }

In this section, we analyze how Russian state media has changed its utilization of Telegram since the beginning of the Russo-Ukrainian War, as well as how the content promoted on the largely uncensored Telegram platform has differed from the content on news media websites. To do so, we first determine whether Russian state media websites promoted Telegram more often than before the onset of the war. Finally, we utilize our topic clustering methodology to determine the topics that were (1)~shared, (2)~specific to Russian media websites, and (3)~specific to Telegram.

\subsection{The Increased Use of Telegram and the Decline of Western Platforms } As seen in Figure~\ref{fig:number-telegrams}, following the Russian invasion of Ukraine, activity within our set of 732~Telegram channels spiked heavily, growing from an average 6,500~messages per day to nearly 27,500. While by September 2022 posting decreased, the average remained near 11,000~messages/day, illustrating a dramatic increase in activity compared to early 2022.  Plotting the top~10 external domains linked to by our set of Russian websites' articles in Figure~\ref{fig:hyperlinks-websites} between January and September 2022, we observe an increase in the promotion of Telegram (t.me domain) by these outlets. While at the beginning of 2022, only 4.0\% of links to external domains were to Telegram, this soon spiked to nearly 10.5\%, remaining above 6.0\% throughout the period measured. This confirms that across Russian media sites, there \textbf{has been a dramatic increase in the use and promotion of Telegram since the beginning of the conflict.}

\begin{table}
\centering
\fontsize{9pt}{8.5pt}
\selectfont
\begin{tabular}{lr}
Top YouTube & \# Telegram  \\
 Channels & Channels  \\ \midrule
\foreignlanguage{russian}{Россия 24} & 61 \\
\foreignlanguage{russian}{Анатолий Шарий} & 50 \\
Max Chronicler & 46  \\
\foreignlanguage{russian}{Комсомольская Правда }  & 39 \\
\foreignlanguage{russian}{Информационное агентство БелТА}  & 39 \\
\foreignlanguage{russian}{RT на русском}  & 34 \\
Metametrica  & 34  \\
Fox News  & 32 \\
\foreignlanguage{russian}{ИЗОЛЕНТА live}   & 31 \\
\foreignlanguage{russian}{Телеканал Рада} & 31 \\
SHAMAN & 29 \\
\end{tabular}
 \caption{\label{tab:top-youtube}Top YouTube channels mentioned by our set of 732~Telegram channels. }  
\vspace{-5pt}
\end{table}

Consistent with both the banning of many Russian media firms by Western social media sites and the corresponding banning of Western social media platforms by the Russian government~\cite{Chee2022, Bergengruen2022}, in Figure~\ref{fig:hyperlinks-websites}, we observe a slight decrease in hyperlinks to Twitter from Russian state media sites. Extracting the external hyperlinks from our set of 732~Telegram channels, we similarly observe a steady decrease in the use of YouTube. Many of the most hyperlinked YouTube channels (Table~\ref{tab:top-youtube}) were pro-Russian channels that were later restricted~\cite{Wong2022}. {We note the presence of several conservative news websites within the top-linked YouTube channels. As noted elsewhere~\cite{Stone2022}, US-based conservatives have sometimes adopted Russian state narratives surrounding the war. This is largely reflected in the number of Telegram channels linking to YouTube channels like Fox News and Metamerica (Table~\ref{tab:top-youtube}).}
We thus see \textbf{a decline in the use of Twitter and YouTube by Russian state media organizations following the invasion of Ukraine}.

\subsection{The Shared Ecosystem\label{sec:similarities-analysis}}

Having observed a noted increase in the usage of Telegram, we now determine the shared and differing content within these different platforms. To understand the degree to which different Russian websites in our dataset conformed to the messages and topics present within Telegram, we determine the semantic similarity between our set of state media websites and linked Telegram channels. Altogether 1,812,648 out of the 2,477,564 (73.2\%) Telegram messages had a corresponding paragraph on a Russian site. Conversely, 899,589 of the 1,616,946 paragraphs (55.6\%) from our set of Russian websites had a corresponding Telegram message. We thus observe a high degree of but not complete similarity in terms of topics on Russian media articles compared to Telegram.

\begin{figure}
\begin{subfigure}{.48\textwidth}
  \centering
  \includegraphics[width=1\linewidth]{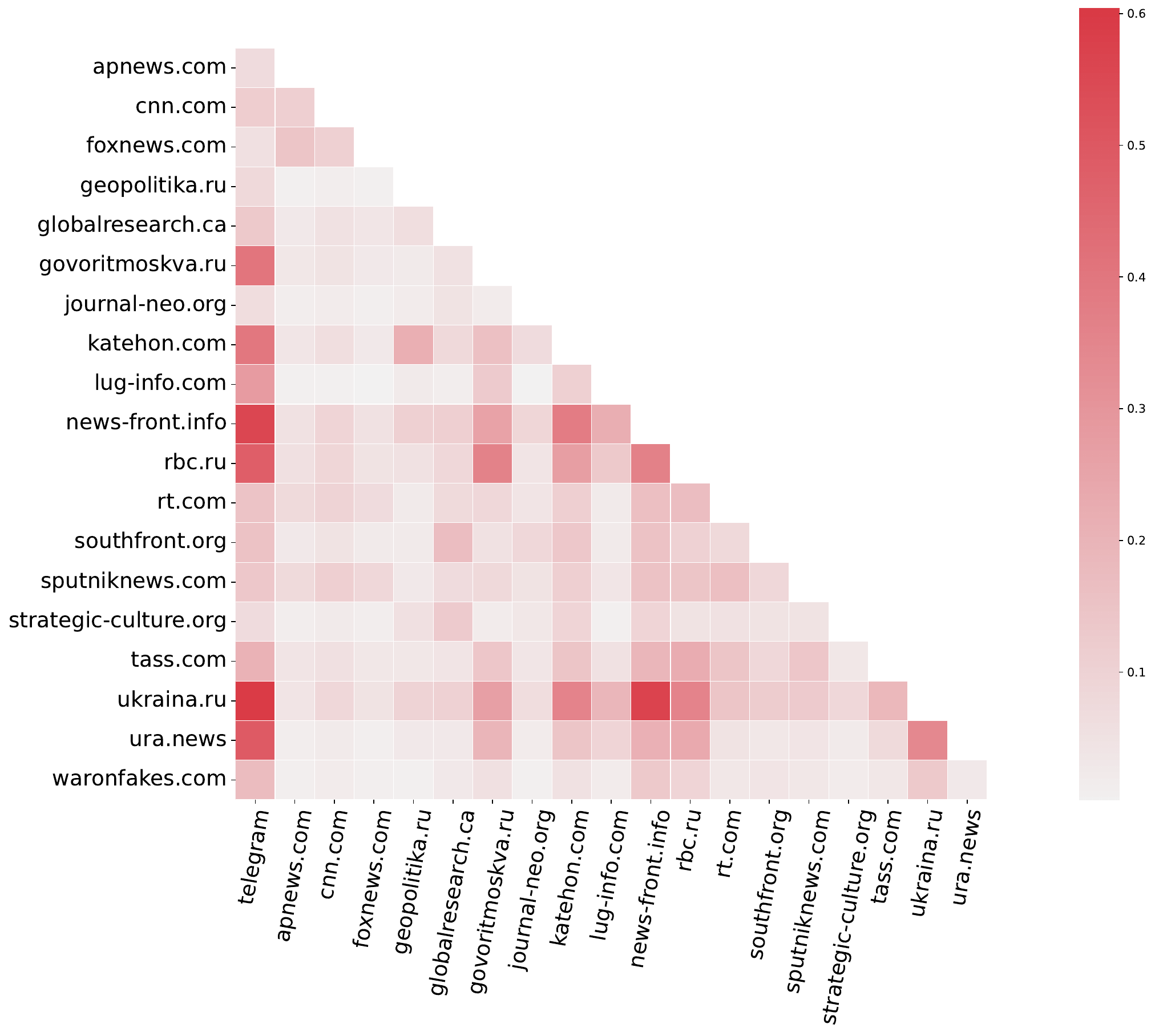}
 
\end{subfigure}%
 \caption{Similarity matrix between our considered websites. {Ukraina.ru} and {newfront.info} have the highest similarity with one another. All Russian websites have a high content similarity with the collective messages posted to our set of 732 Telegrams. The three US-based websites have high similarities amongst themselves.
}
\label{fig:similarity_matrix}
\vspace{-15pt}
\end{figure}

\vspace{2pt}
\noindent
\textit{\textbf{Similarity to Telegram.}} We now determine the similarity scores between our given platforms (Section~\ref{sec:similiarty-scores}) based on the geometric mean of their percentages of shared messages. We note that for this analysis, we combine our set of Telegram messages (as if all came from one website), rather than examining each of the Telegram channels individually. To provide a reference point, as well as to validate our metric, we scrape and compare the similarities for three English-language news websites: {cnn.com}, {foxnews.com}, and  {apnews.com}. Using the methodology outlined in Section~\ref{sec:data-eng}, we gather an additional 41,452, 78,494, and 104,206 articles from {cnn.com}, {foxnews.com}, and  {apnews.com} respectively. We follow the same methodology outlined for segmenting and embedding the articles (Section~\ref{sec:preprocessing}). 

Calculating the similarities between each website and Telegram, as seen in Figure~\ref{fig:similarity_matrix}, the US-based websites have the highest semantic similarity with each other, and the second highest semantic similarities with the Russian websites {sputniknews.com}, {rt.com}, and {tass.com}. We note that all of the Russian websites included in our dataset have relatively high semantic similarity with our set of Telegram messages compared with {cnn.com}, {foxnews.com}, and {apnews.com} with ura.news having the highest similarity and journal-neo.org having the least. 



\begin{table}[t]
\centering
\fontsize{9.0pt}{4.5pt}
\selectfont
\setlength{\tabcolsep}{4pt}
\begin{tabularx}{\columnwidth}{XXr}
 \textbf{Website}& \textbf{Sim. Telegram} & \textbf{Simalarity} \\ \midrule
geopolitika.ru & rossiyaneevropa & 0.116 \\
globalresearch.ca & gr-crg & 0.109 \\
govoritmoskva.ru & radiogovoritmsk & 0.340
 \\
journal-neo.org & rossiyaneevropa
 &  0.095\\

katehon.com & rossiyaneevropa &  0.364 \\
lug-info.com & lic-lpr&  0.516 \\
news-front.info &ukraina\_ru& 0.473
 \\
rbc.ru & rbc\_news &  0.486\\
rt.com & vzglyad-ru&  0.135 \\
southfront.org &southfronteng &   0.150 \\
sputniknews.com & vzglyad\_ru &  0.125\\
strategic-culture.org & strategic\_culture &0.087\\
tass.com & tassagency\_en & 0.273 \\
ukraina.ru & ukraina\_ru & 0.526 \\
ura.news &  vzglyad\_ru&  0.354 \\
waronfakes.com & warfakes&0.299 \\
\end{tabularx}
\caption{\label{tab:news-websites-to-telegrams} Most similar Telegram chanels to each website.} 
\end{table}


\begin{figure}
  \centering
  \includegraphics[width=0.5\linewidth]{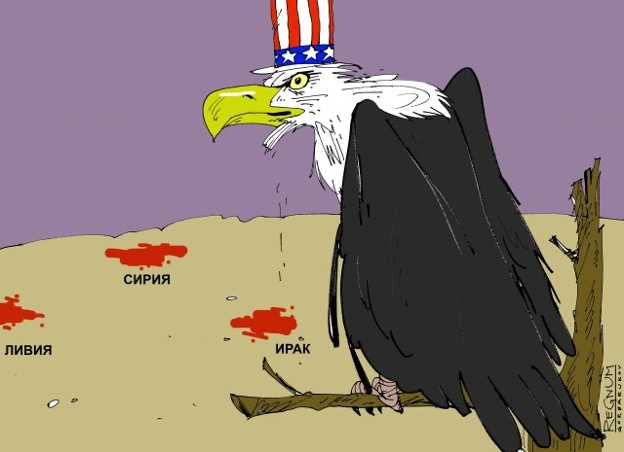}
  

\scriptsize{
 \caption{Propaganda image from @rossiyaneevropa. @rossiyaneevropa argues NATO cannot criticize Russia's activities in Ukraine given the West's war crimes in Libya.\label{fig:propaganda-image}}
 }%
\vspace{-10pt}
\end{figure}

\begin{table*}
\centering
\fontsize{9.0pt}{3.5pt}
\selectfont
\setlength{\tabcolsep}{2pt}
\begin{tabularx}{\textwidth}{l|XrXrXr}
 &\multicolumn{1}{c}{{\textbf{\scriptsize{Telegram}} }} &   &\multicolumn{1}{c}{{\textbf{{Russian Site }}}} &  & \multicolumn{1}{c}{{\textbf{{Shared Telegram} }} } & Telegrams\\ 
&  \multicolumn{1}{c}{{\textbf{{Specific Topics}} }} &  Telegrams  & \multicolumn{1}{c}{{\textbf{{Specific Topics}} }}  & Articles &  \multicolumn{1}{c}{{\textbf{{and Russian Site Topics}} }}  & +Articles \\\midrule
1 & Lukashenka on the possibility of achieving peace in Donbas (Russian $\rightarrow$ English)& 5,295 & According to him, the problem is very acute, especially since now it is very difficult to engage in unauthorized extraction of gas going to Europe from the Ukrainian pipeline because this will be monitored not only in Russia, which is losing money but also in Europe. (Russian $\rightarrow$ English) &635 & Regarding the events in Shchastya, Luhansk region, where militants hit the building of the fire and rescue department (Ukrainian $\rightarrow$ English) & 19,467 \\ 
2& Footage of live firing of Grad multiple launch rocket systems of a self-propelled artillery regiment of the Central Military District in the zone of a special military operation. (Russian $\rightarrow$ English)& 2,326  
& Mass riots in Iran get out of control of the security forces Amid the riots in Iran, caused, allegedly, by direct incitement from the United States (Russian $\rightarrow$ English) & 436 & The Russian Defense Ministry showed the advancement of airborne units during a special military operation in Ukraine. (Russian $\rightarrow$ English) & 9,939\\
3& Krajina security forces will not be given a corridor to exit Mariupol, Basurin said. (Russian $\rightarrow$ English)& 2,182 & Cessation of hostilities and political dialogue, negotiations, mediation, and other peaceful means aimed at achieving a lasting peace. (Russian $\rightarrow$ English)& 197 & Today, the Russian military attacked the Khmelnytsky region from the air. According to the head of the Khmelnytsky OVA Serhiy Gamaly, there were no casualties. (Ukrainian $\rightarrow$ English) & 5,217\\

\end{tabularx}

\caption{\label{tab:shared-disinct-content} The top three topics---by the number of distinct articles or distinct Telegram messages---within the Telegram-specific, the Russian website-specific, and the shared Telegram and Russian website ecosystems.  }  
\vspace{-10pt}
\end{table*}

\vspace{2pt}
\noindent 
\textit{\textbf{Most Similar Telegram Channels to Russian Websites.}} To further examine the correspondence between Telegram and our set of Russian media outlets, we determine the most similar sets of Telegram channels to each website. As seen in Table~\ref{tab:news-websites-to-telegrams}, several Telegram channels individually post many of the same topics present on Russian websites. Unsurprisingly, across our set of websites, the most similar Telegram channels to each of our websites are the official Telegram channels utilized by these online newspapers. 

Besides the official channels, one of the the channels most similar to our Russian websites is {@rossiyaneevropa} (14.6K~subscribers), a pro-Russian channel, ostensibly run by ``Alexander Burenkov, director of the Institute of Russian-Slavonic Studies.'' As pictured in Figure~\ref{fig:propaganda-image}, this channel often amplifies Russian propaganda. We further observe that several other similar Telegram channels are run by Russian state-controlled media (@vzglyad\_ru, 88.1K subscribers) or by Russian-backed Ukrainian-separatists (@nm\_dnr, 85.5K subscribers; @lic\_lpr 31.3K subscribers; @gtrklnr\_lugansk24 7.4K; @dnr\_sckk 17.2K). In addition, to these channels, we further find two of the other most similar Telegrams (not shown in Table~\ref{tab:news-websites-to-telegrams}) {@rentv\_news} (49.7K subscribers) and {@killnet\_reservs} (93.9K subscribers) are pro-Russian channels that often repeat or rebroadcast videos, official documents from the Russian government, disinformation, and anti-American and anti-Ukrainian messages. For example, on October 10, 2022, {@killnet\_reservs} posted a message advocating for their subscribers to disrupt US infrastructure: 

\begin{displayquote}
\small
\textit{We invite everyone to commit DDOS on the civilian network infrastructure of the United States of America! Subject to DDOS attacks are:- All Airports - Marine terminals and logistics facilities - Monitoring weather centers - Health care system - Metro (buying tickets, registering the route) - Exchanges and online trading systems.}\footnote{\url{https://web.archive.org/web/20221011223806/https://t.me/s/killnet_reservs}}

\end{displayquote}

We note that while we utilize this methodology to identify the messages that bear the closest resemblance to specific Russian websites, and thus to content sponsored by Russian=-state media, among our set of 732~Telegram channels, this methodology can be easily extended to identify suspect Telegram channels that repeat or mention propaganda on a much wider scale, which we leave to future work.  


\vspace{2pt}
\noindent 
\textit{\textbf{Most Prominent Topics in Shared Ecosystems.}} Next, to understand the most prominent topics present among all shared texts between Telegram and Russian websites, we utilize our clustering algorithm outlined in Section~\ref{sec:cluster}. Specifically, we cluster only the set of paragraphs from Russian websites that had a corresponding Telegram message together with all Telegram messages that had a corresponding paragraph on a Russian website; altogether this consists of 2.7M~messages/paragraphs (66.24\% of all texts [from both Telegram and Russian articles]). We note, as previously discussed, that due to the intrinsic guarantees of our clustering algorithms, each embedding has a high cosine similarity with the center. After clustering with $\lambda=0.8$, our embeddings had an average of 0.882 cosine similarity with their respective cluster centers. Each cluster had an average cosine similarity of 0.0292 with the remaining cluster centers. This illustrates the degree of semantic cohesiveness within each cluster and that each cluster is distinct. For each cluster, to build an intuition for the topic, we extract the most representative paragraph/message from the cluster by getting the paragraph with the highest cosine similarity to the cluster center. 


Determining the top topics by the number of articles, the most shared topic within our datasets concerned the destruction of buildings following Russia's invasion of Ukraine, specifically, the operations in the Luhansk region (Table~\ref{tab:shared-disinct-content}). This was largely anticipated given how pivotal the Donbas region became to the war and propaganda surrounding it~\cite{laryvs2022delegated}; 19,467 separate articles (note, not paragraphs) and Telegram messages mention the topic. The second most popular topic with 9,939~messages/articles was about the actual invasion of Ukraine by Russian units. Within Vladimir Putin's declaration indicating Russia's intention to invade Ukraine, he called for a ``special military operation'' to ``denazify'' and ``demilitarize'' Ukraine. We see this particular language repeated by both Russian propaganda outlets and Telegram channels. The third topic (again a specific aspect of the war) concerns the Russian bombing of the Khmelnytsky region of Ukraine on April 28, 2022.



\subsection{Topics Specific to Russian News Websites}
Having explored the shared content amongst Russian news and Telegram channels, we now analyze the content specific to our set of Russian news websites. We cluster only the set of paragraphs from Russian sites that did not have a corresponding Telegram message in our dataset (727,357~paragraphs from 120,198~articles). Across this clustering, each paragraph had an average 0.926~cosine similarity with their respective cluster center. Each cluster has an average similarity of 0.0674~with other clusters in the dataset.


As seen in Table~\ref{tab:shared-disinct-content}, the most common topic specific to Russian news websites concerned the extraction of gas from Ukraine to Europe; an estimated 635~articles (note, not paragraphs) across our 16~news websites were associated with this topic. As noted by others, following the Russian invasion, the sanctions on Russian gas and oil to Europe have caused enormous major economic fallout~\cite{Soldatkin2022} and we see mentions of this topic repeated across our Russian media ecosystem. Besides content concerning the shipment of gas from Ukraine to Europe, we further see content about protests in Iran that followed the death of Mahsa Amini on September 16, 2022 (436~articles) and calls for peace talks between Russia and Ukraine (197~articles).

\subsection{Topics Specific to Telegram}
We now extract content specific to Telegram. To do so, we cluster only the set of messages from our Telegram channels that did not have a corresponding text from our Russian media outlets' articles. Altogether, we cluster the 664,916~distinct Telegram messages. Across this clustering, each Telegram message had an average 0.900~cosine similarity with their respective cluster center. Each cluster center has an average cosine similarity of -0.0708 with the other clusters. 

As seen in Table~\ref{tab:shared-disinct-content}, many of the most prominent messages concern specific day-to-day updates on the war in Ukraine (Topic~2 and~3). For example, the topic cluster with the second most Telegram messages concerns updates on the rocket launches of a Ukrainian city~\cite{Holder2022}. The Telegram topic with the most corresponding messages however concerns an interview with Belarusian President Alexander Lukashenko on the possibility of peace in the Donbas region of Ukraine. We thus see within many of these Telegram-specific content clusters highly specific updates about news and individual interviews rather than larger news stories.


\section{The Spread of Topics}
Having analyzed the static behavior of the topics shared among our Telegram and Russian news ecosystems, in this section, we examine the speed at which topics spread amongst and between Russian websites and Telegram channels. To estimate the spread of topics amongst and between different Russian websites and Telegram channels, we first cluster {all} 4,094,510 texts (Telegram messages and paragraphs from articles) as specified in Section~\ref{sec:cluster}. Across this new clustering, with 29,875~unique centers, each message/paragraph had an average cosine similarity of 0.883, with its respective cluster center. Each cluster further had an average cosine similarity of 0.0321 with the remaining clusters. We note that when performing our analysis using the timestamps of our articles and Telegram messages given that we only managed to determine the publish date and not the \emph{exact} timing information of our articles' publication (\textit{i.e.}, the hour and second of publication), we perform our analysis on the scale of days. This is such that we consider a Telegram message to {precede an article only if it was published at least a day before an article (\textit{e.g.}, 04-05-2022 before 04-06-2022}).
\begin{table}
\centering
\fontsize{9.0pt}{2.5pt}
\selectfont
\setlength{\tabcolsep}{2pt}
\begin{tabular}{lrlr }
 &\%  First &  &  \%  First\\
Website & on Telegram & Website & on Telegram \\  \midrule
geopolitika.ru &  {8.10}\% &   rt.com &  6.70\%\\
globalresearch.ca & 4.60\% &southfront.org &8.55\%\\
govoritmoskva.ru & 11.6\% & sputniknews.com & 5.06\% \\
journal-neo.org &  7.83\% & strategic-culture.org & 6.06\% \\
katehon.com &15.0\%  & tass.com & 11.6\% \\
lug-info.com &  19.7\% & ukraina.ru & 27.9\%\\
news-front.info & {21.2\%} & ura.news & {25.6\%} \\
rbc.ru & 16.1\%& waronfakes.com &  \textbf{28.2\%} \\
\end{tabular}
\caption{\label{tab:began-on-telegram} Percentage of each website's topics that were posted first on Telegram (after removal of official channels). We bold the largest percentage.}  
\end{table}


\begin{table}[t]
\centering
\fontsize{9.0pt}{4.0pt}
\selectfont
\setlength{\tabcolsep}{3pt}
\begin{tabularx}{\columnwidth}{XXr}
 &  & \% Content \\ 
Website& Telegram &First Posted \\

\midrule
geopolitika.ru & rossiyaneevropa& 0.47\%\\
globalresearch.ca & karaulny & 0.32\% \\ 
govoritmoskva.ru & karaulny& 1.5\%
 \\
journal-neo.org & parstodayrussian
 &  0.50\% \\

katehon.com & karaulny&  1.0\%\\
lug-info.com & lic\_lpr &  2.7\% \\
news-front.info &karaulny & 1.2\%\\
rbc.ru & karaulny &  1.6\%\\
rt.com & karaulny&  0.43\% \\
southfront.org &new\_militarycolumnist &   0.57\%\\
sputniknews.com & karaulny & 0.38\%\\
strategic-culture.org & karaulny &0.37\%\\
tass.com & karaulny& 0.61\%\\midrule
ukraina.ru & karaulny & 1.7\%\\
ura.news &  karaulny &  2.6\% \\
waronfakes.com & senkevichonline &5.5\%
 \\

\end{tabularx}
\caption{\label{tab:news-websites-to-telegrams2} Top telegrams --- by percentage of content first posted --- that post content prior to our set of websites.}  
\end{table}
\noindent 
\textbf{\textit{Content First Published on Telegram and Russian Websites.}} To understand the interchange of topics between our Russian website and Telegram datasets, we first determine the percentage of each website's topics that began on Telegram and \textit{vice versa}. We note that for this analysis, we remove the set of Telegram channels that officially operate in coordination with each of our news websites. This enables us to determine how much each website's content began on ostensibly ``independent'' (\textit{i.e.}, not controlled by same Rusian-state media entities) Telegram channels.\footnote{We remove the following Telegram channels: tass\_agency, tassagency\_en, uranews, radiogovoritmsk, ukraina\_ru, rbc\_news, southfronteng, strategic\_culture, waronfakesen, warfakeres, warfakes, warfakebelgorod, warfakeskrm, warfakeszo, and warfakers.}  Determining the amount of a website's content that each cluster contains (\textit{i.e.}, the number of paragraphs), we thus determine the percentage of each platform's content that was first posted on Telegram (Table~\ref{tab:began-on-telegram}).
 

As seen in Table~\ref{tab:began-on-telegram}, even after removing official Russian Telegram channels, most of our websites had a noticeable portion of their content that was first posted on Telegram. This is particularly true of websites like {waronfakes.com} (28.2\%), {ura.news} (25.6\%), and {ukraina.ru} (27.9\%). Across all websites, 13.9\% of their topics began on Telegram, accounting for 18.4\% of all paragraphs in our dataset. We further see that as a whole 33.2\% of Telegram's topics began on Russian websites, accounting for 24.3\% of all Telegram content. Thus while several of our news websites like waronfakes.com, ura.news, ukraina.ru, and rbc.ru, indeed publish content after it first appears (at least a day later) on Telegram, Telegram channels themselves also utilize topics from Russian news sources for nearly a third of their topics.

 Examining the set of Telegram channels that most often first posted each website's content in Table~\ref{tab:news-websites-to-telegrams2}, we see the Telegram channels @kalaruny (159K subscribers), @new\_militarycolumnist (213K), and @bbbreaking (1.38M; not shown in table) often posted content first across nearly all of our websites. As reported elsewhere, @karaulny (screen name Karaulny-Z) is a pro-Russian government channel that abides by a ``stop list'' of prohibited topics and that was purchased by supporters of the Kremlin in 2017~\cite{Rothrock2018}. @new\_militarycolumnist, another pro-Russian channel, gives continuous updates on the Russian military~\cite{Sabah2022}. Finally, @bbbreaking, another pro-Russian Russian-language channel, with the motto \textit{``Earlier than others. Almost''} appears to also in several cases give updates on the news before many of our Russian news websites. 

%

\vspace{2pt}
\noindent
\textbf{\textit{Influence Estimation through Hawkes Processes.}} As just seen, a substantial portion of the content on websites like waronfakes.com, ura.news, and rbc.ru first appears on Telegram. Despite this and even given the close relationship that many of these websites have with our set of Telegram channels, it is largely infeasible to \emph{know} definitively if these topics' appearance on Telegram \emph{caused} their later appearance on our set of websites. However, in this section, we utilize Hawkes processes, to estimate the probability that a message first appearing on a given Telegram influenced a given website to write about similar content. Utilizing this approach, we give the estimated percentage of content that appears on one platform that may have been influenced by another platform as well as the efficiency of this influence. 
\begin{figure}[t]
\begin{subfigure}{.45\textwidth}
  \centering
\includegraphics[width=1\linewidth]{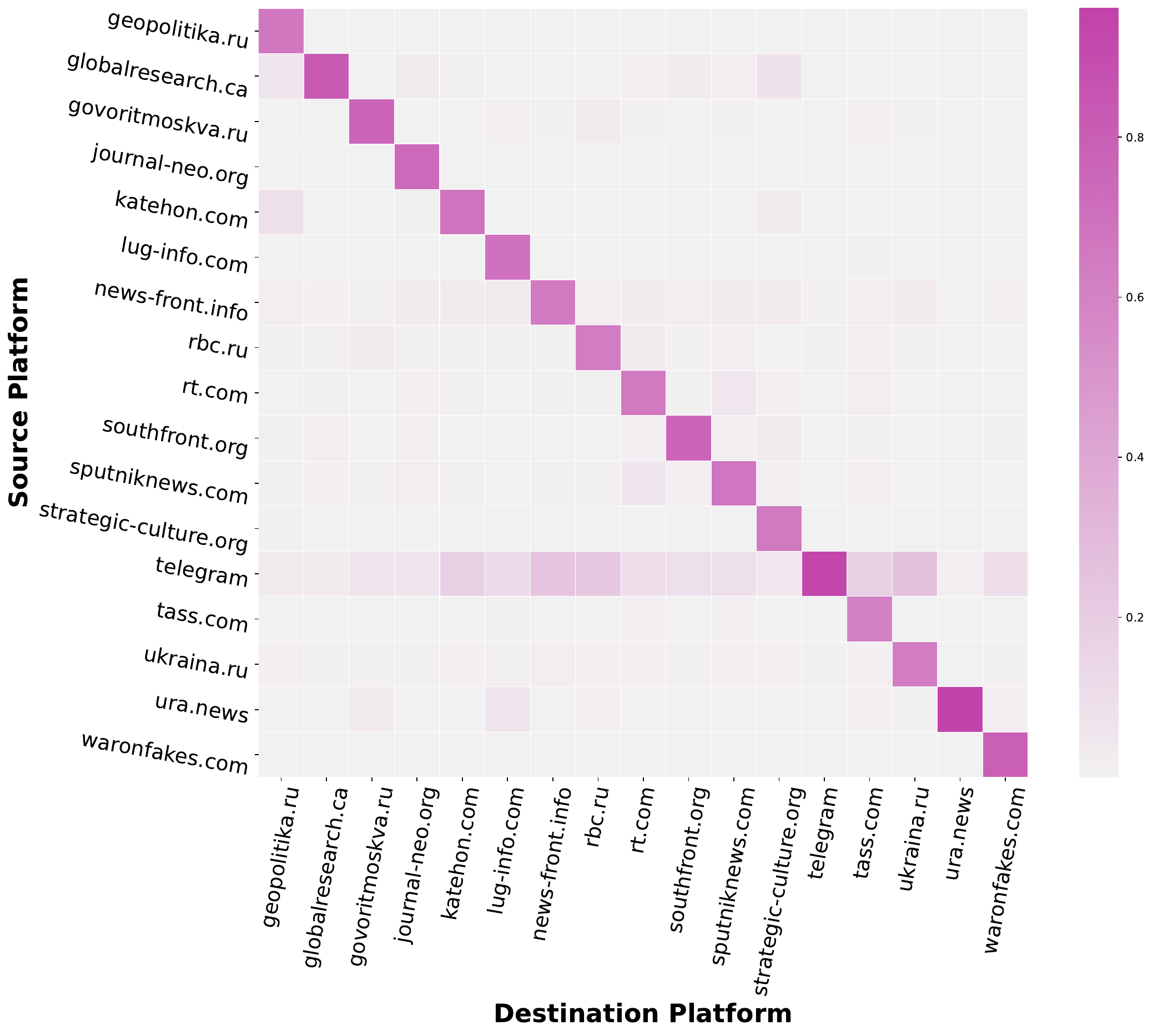}
\end{subfigure}%
\caption{ We report the percentage of each platform's content that was estimated (using the methodology outlined in Section~\ref{sec:methoddology}) to have been influenced by another platform.}
\label{fig:hakews}
\vspace{-12pt}
\end{figure}

\begin{figure}[t]
\begin{subfigure}{.45\textwidth}
  \centering
\includegraphics[width=1\linewidth]{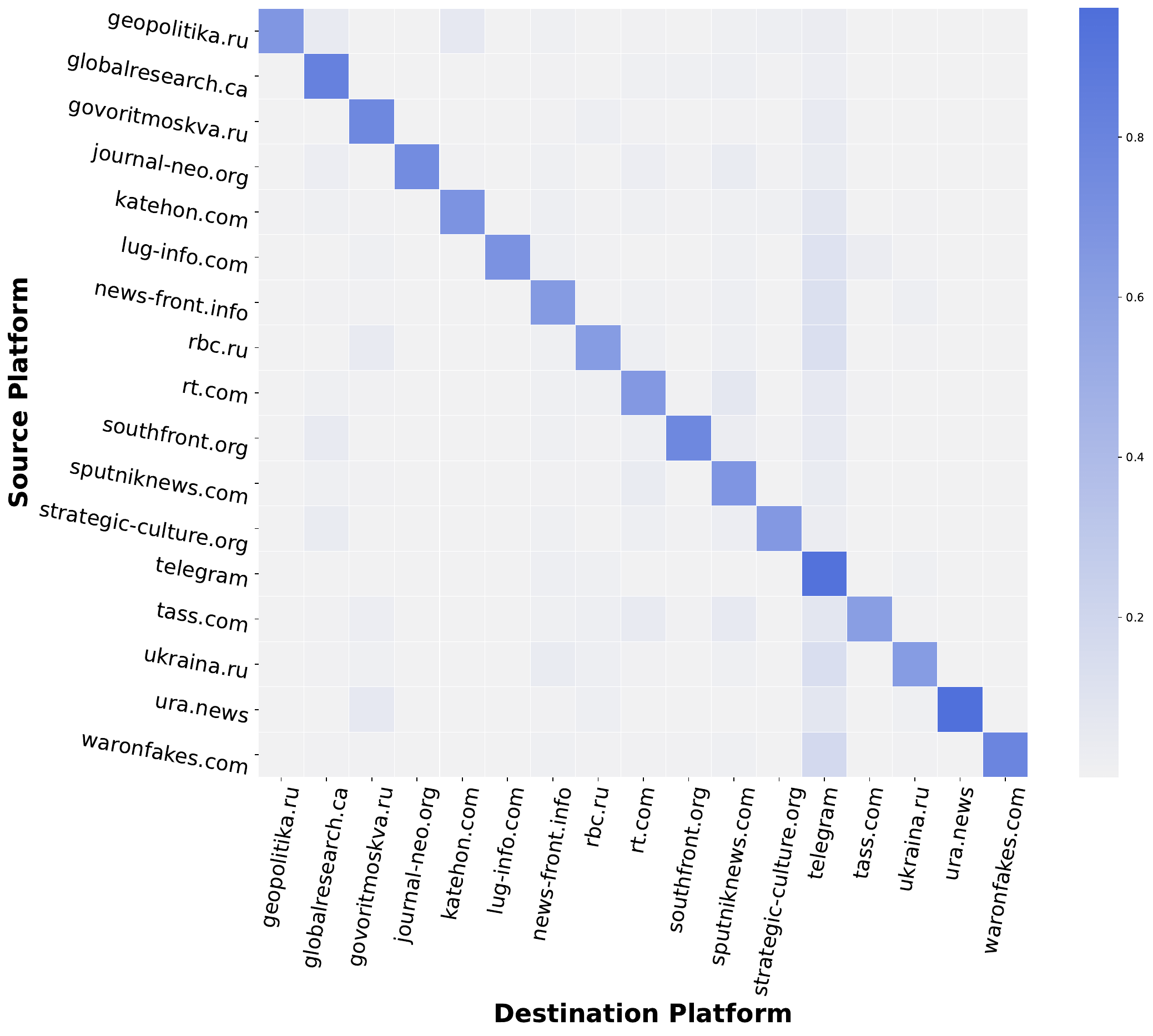}
\end{subfigure}%
\caption{ We report the estimated efficiency (using the methodology outlined in Section~\ref{sec:methoddology}) of each platform in getting their content onto different platforms.}
\label{fig:hakews-efficiency}
\vspace{-12pt}
\end{figure}

 As previously specified, to estimate the influence of one platform on another (Section~\ref{sec:background}), we fit 17 Hawkes interacting processes (one for each platform) using Gibbs sampling for our 29.9K topic clusters and utilizing the daily frequencies of each website reporting on each of these topics~\cite{linderman2015scalable,zannettou2018origins}. We report the estimated percentage of each platform's paragraph/messages that may have been caused by the other platforms in our dataset (Figure~\ref{fig:hakews}). We note that because we limited our study to our set of 16~websites and 732~Telegram channels, other platforms' influence would be included within each website's estimated self-influence as its background rate.

\begin{figure*}
\begin{subfigure}{.5\textwidth}
  \centering
  \includegraphics[width=1\linewidth]{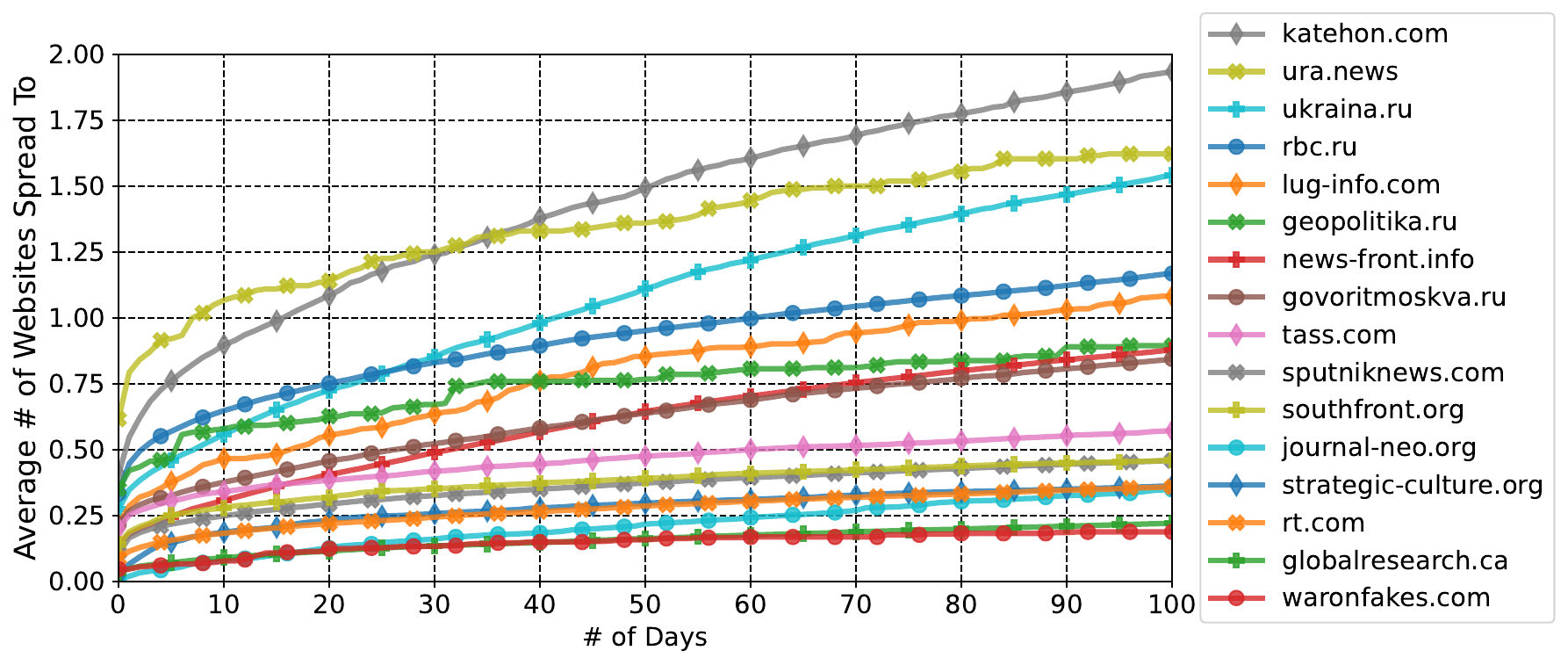}
    \vspace{-8pt}
  \caption{Russian Websites to Russian Websites}
\label{fig:russian-website-to-website}
\end{subfigure}%
\begin{subfigure}{.5\textwidth}
  \centering
  \includegraphics[width=1\linewidth]{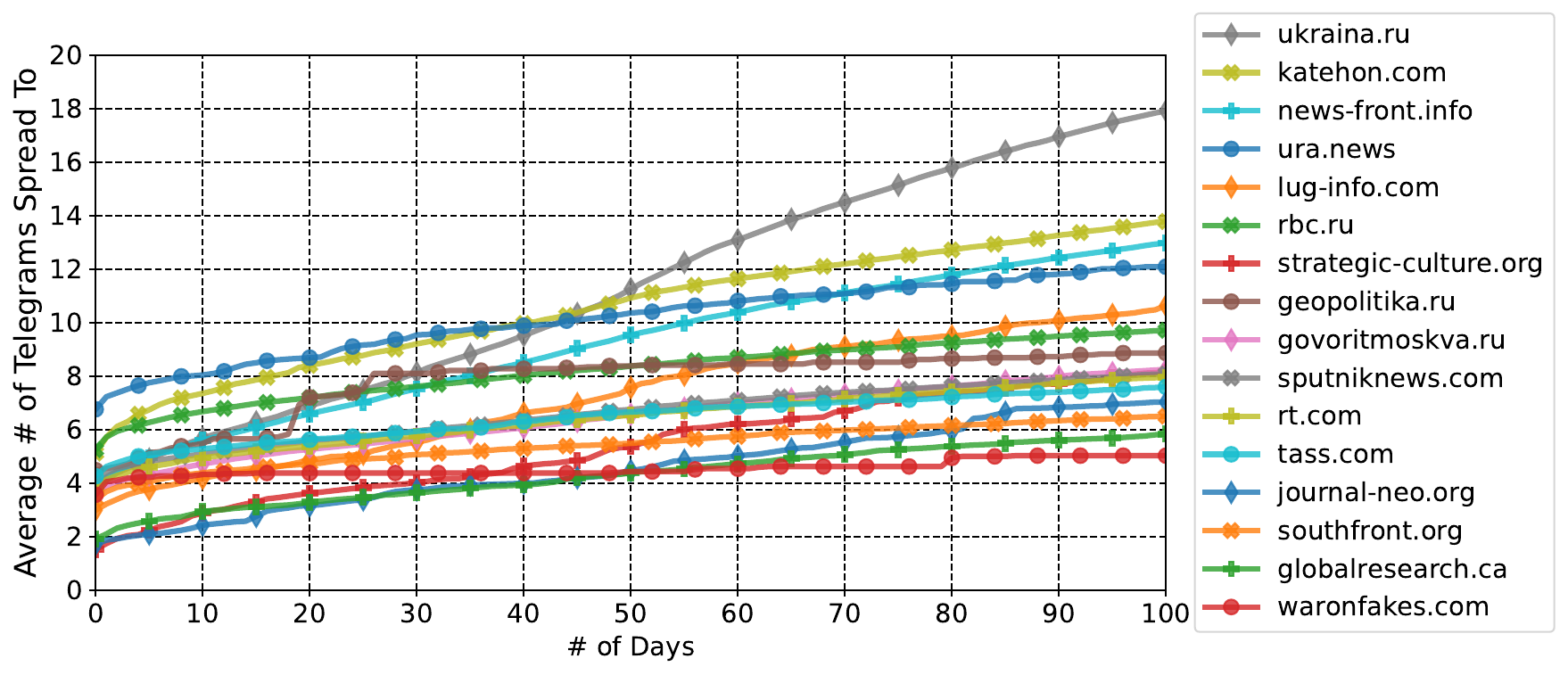}
    \vspace{-8pt}
  \caption{Russian Websites to Telegram}
\label{fig:russian-website-to-telegram}
\end{subfigure}%
\caption{Katehon.com is the most effective at getting its original content reposted by other Russian news websites. Despite many Russian websites utilizing Telegram, it often takes months on average before a topic is widely addressed by more than a few dozen channels on Telegram.
}
\vspace{-15pt}
\end{figure*}
\begin{figure*}
\begin{subfigure}{.49\textwidth}
  \centering
  \includegraphics[width=1\linewidth]{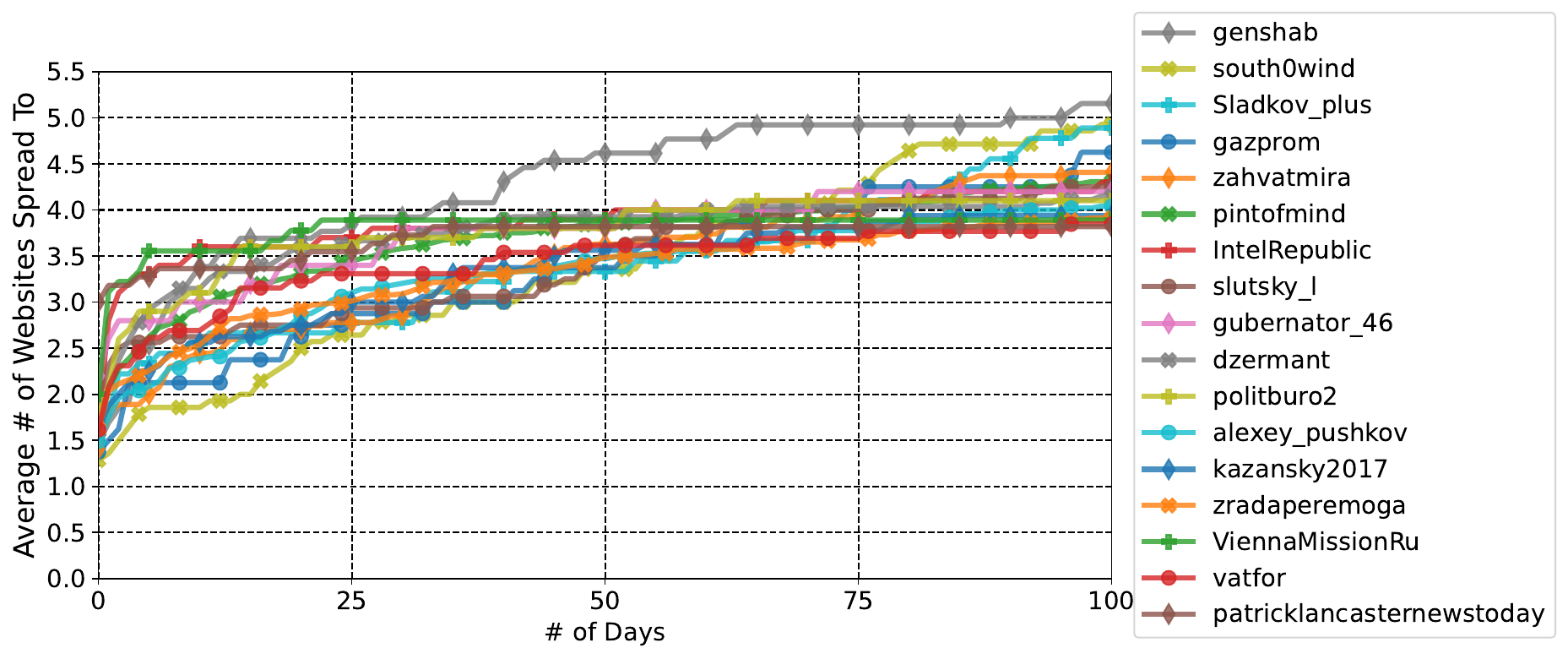}
  \caption{Telegram to Russian Websites}
\label{fig:from-telegram-totelegram}
\end{subfigure}%
\begin{subfigure}{.49\textwidth}
  \centering
  \includegraphics[width=1\linewidth]{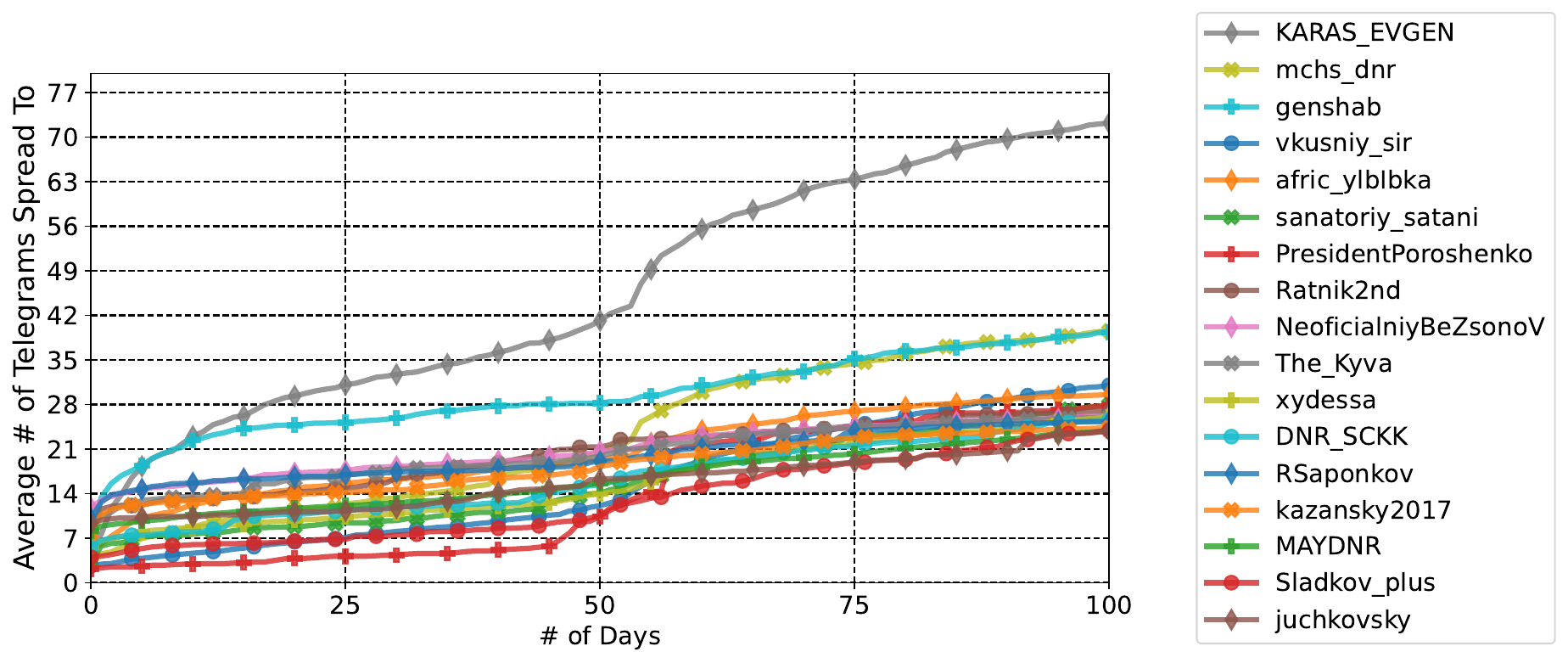}
  \caption{Telegram to Telegram}
\label{fig:from-telegram-to-websites}
\end{subfigure}%
\caption{ Despite their smaller size several Telegram channels post topics and content that are later repeated by Russian news outlets. The pro-Russian channel @{genshab} is particularly effective at getting its content echoed within this ecosystem.
}
\label{fig:narratives-from-telegram}
\vspace{-10pt}
\end{figure*}
Plotting the estimated influence of each website on each other and Telegram, we see in Figure~\ref{fig:hakews}, that Telegram typically has a moderate estimated influence on the content published on each of the websites. Most prominently, Telegram has a possible influence on 18.4\% of content on katehon.com, 25.2\% on news-front.info, 23.0\% on rbc.ru, and 26.7\% on ukraina.ru. Conversely, it has the weakest influence on the website ura.news (despite often writing about its topics prior to ura.news [Table~\ref{tab:began-on-telegram}]). With the smallest amount of articles, we see that geopolitika.ru also had the smallest estimated influence on Telegram while news-front.info had the greatest effect on Telegram. Further as largely expected, we see in the estimated topic spread efficiency in Figure~\ref{fig:hakews-efficiency} that many of our websites are somewhat efficient at seeing their content migrate to Telegram, with waronfakes.com being the best at 18.8\% efficiency. This indicates across our websites, many may have to post a few articles before that article's topic appears on Telegram.


In addition, to estimating the influence of each website on our set of Telegram channels, we further estimate the relationships that each website has with one another. As largely expected prominent websites like rt.com and sputniknews.com have the largest estimated possible influence on the content of one another relative to other websites (around 6\% [Figure~\ref{fig:hakews}]). Looking at the efficiency of these influence relationships, we see again a stronger relationship between many of the large English-language Russian news sites (tass.com, sputniknews.com, and rt.com). Similarly, we see again that news-front.info, which has been previously documented as heavily influencing conversations within Russian propaganda ecosystems~\cite{hanley2022happenstance} has a marked estimated influence on other websites, compared to the rest of the websites, playing a possible role in nearly 3\% of the content on each of other websites (Figure~\ref{fig:hakews}).

\noindent
\textbf{\textit{Speed of Spread of Topics From Russian Websites.}} Having estimated the influence of each website, we now model and determine how quickly, on average, \textit{original content} (\textit{i.e.}, content whose topic was first posted on a particular website or a Telegram channel) from each of our websites and Telegram channels travel to \textit{other} websites and Telegram channels. As in Section~\ref{sec:similarities-analysis}, we perform this analysis on the scale of days. As seen in Figure~\ref{fig:russian-website-to-website}, on average, {katehon.com} and ura.news's topics tend to spread the fastest. Across all topics, we further observe that after initially spiking in popularity on the first day it is published, topics often take a longer period to spread throughout the entire ecosystem. Examining each topic individually, we found one of the fastest topics to spread was a story that originated on sputniknews.com, rt.com, and tass.com. This story, which reached every website within six days, was about a potential United Nations-brokered meeting between Russian Federation President Vladimir Putin and Ukrainian President Volodymyr Zelenskyy.\footnote{\url{https://web.archive.org/web/20220919021146/https://sputniknews.com/20220919/putin-zelensky-meeting-far-from-possible-but-un-ready-to-help-facilitate-guterres-says-1100939551.html}}

Examining the spread of each website's content to our set of 732~Telegram channels in Figure~\ref{fig:russian-website-to-telegram}, we see that on average, stories did not typically spread to more than approximately 20~Telegram channels within the first 100~days. Furthermore, we see that the websites whose content spreads the fastest {katehon.com} and {ura.news}, are again the websites whose content spreads the quickest on Telegram. Indeed, we see among our websites, {katehon.com}, {ura.news}, ukraina.ru, and news-front.info, are the best progenitors of content across both other websites as well as Telegram. This adheres to results in prior work that show {newsfront.info}  and {katehon.com} as two of the key creators of Russian propaganda on the Internet~\cite{hanley2022happenstance,RussiaPillar2020}. Examining one of the most prolific stories, one originating from {katehon.com} published the military results of the first day of the Russian invasion of Ukraine and spread to 613 different Telegram channels.\footnote{\url{https://web.archive.org/web/20220510150650/https://katehon.com/ru/news/igor-strelkov-itogi-pervogo-dnya-boevyh-deystviy-i-ih-kratkiy-analiz}} 

\vspace{2pt}
\noindent
\textbf{\textit{Spread of Topics From Telegram Channels.}} We now examine the rate at which topics spread from some Telegram channels amongst themselves \emph{and} to Russian websites. We display the set of 16 Telegram channels whose content spreads the furthest (\textit{i.e.}, the channels that originated topics that then spread prolifically and widely) in Figure~\ref{fig:narratives-from-telegram}. As seen in Figure~\ref{fig:narratives-from-telegram}, one of the Telegram channels most effective at originating content/topics that are then reposted elsewhere \emph{both} on Telegram and within the Russian website ecosystem is {@genshab} (86.9K subscribers). A pro-Russian propaganda channel, the channel continuously comments on the Russian invasion of Ukraine, writing on August 29th: 
\begin{displayquote}
\small

\textit{``Kyiv launched a widely publicized ``offensive'' on Kherson.
Due to the clumsy, on the verge of debility fake propaganda, the offensive is developing so far only in the minds of propagandists and those who believe them.''}\footnote{\url{https://web.archive.org/web/20220829120625/https://t.me/genshab/846}}  
\end{displayquote}
Finally, again examining the individual topics from Telegram channels that spread the furthest (the most amount of Telegram channels), we see a message from {@uranews} about the Donbas regions of Ukraine ostensibly wanting to declare independence and join Russia that spread to 557 other Telegram channels.\footnote{\url{https://web.archive.org/web/20220515075654/https://t.me/uranews/41595}}
\begin{displayquote}
\small

\textit{Everyone to defend the Motherland! Donbass is our land!" The DPR authorities, with the help of billboards, are calling on the residents of Donetsk to take arms in their hands to defend the republic from Ukrainian aggression.}\footnote{\url{https://web.archive.org/web/20220515083452/https://t.me/optimisticus007/155}}
\end{displayquote}

\noindent
We further see a message from the pro-Russian Telegram channel~{@optimisticus007} (35.5K~subscribers) that spread to all our websites echoing the desire for Russia to win in Ukraine.
\begin{displayquote}
\small
\textit{Zhirinovsky's last speech in the State Duma: We must win. This is our last decisive battle - the spring of 22. Let it be in April, May, but this year it can be done. We will win!}
\end{displayquote}

\section{Discussion and Conclusion}
In this work, we explored the usage of Telegram by Russian news websites. To do this, we introduced a new, scalable methodology for tracking news narratives across multiple languages and platforms. Our approach, unlike past work, does not depend on dimensionality reduction across all articles at once nor a complete pair-wise similarity calculation, which allows us to scale to several orders of magnitude more articles and to track topics across 215K~news articles and 2.48M~Telegram messages. We find that, unexpectedly, Russian media organizations have not only dramatically increased their usage of Telegram but also are regularly sourcing news topics from the messaging platform. We discuss several implications and future directions of this work here.

\vspace{2pt}
\noindent
\textbf{\textit{Identifying Propaganda Channels.}}
As briefly discussed in Section~\ref{sec:comparing}, our approach can also be used to identify new Telegram channels that align with Russian state-promoted narratives. As was seen in Table~\ref{tab:news-websites-to-telegrams}, our method identified Telegram channels that supported anti-Ukrainian and pro-Russian-separatist sentiments in an automated fashion (\textit{e.g.}, @lic\_lpr, @gtrklnr\_lugansk24, @dnr\_sckk) as well as several channels that largely repeated Russian state-media topics (\textit{e.g.,} @rossiyaneevropa, @pintofmind, @russtrat)~\cite{Afroz2022}. We note that while this method requires lists of predefined propaganda or disinformation platforms/websites, our method can further be utilized beyond Russian and Ukrainian-focused topics to identify websites that are similar to disinformation, hyperpartisan, or other types of malicious websites. 

\vspace{2pt}
\noindent
\textbf{\textit{Content Spreading From Different Platforms.}} By utilizing MPNet, DP-Means clustering, and Hawkes processes, we estimated the role of Telegram and identified the most influential Russian websites within our ecosystem. Unlike past work that has relied on the presence of hyperlinks~\cite{hanley2022no}, our approach approximates influence by directly measuring topics themselves. This approach can be extended to understand how larger platforms like Facebook, Reddit, and Twitter interact with Russian propaganda. While we limited ourselves to analyzing the semantic correspondence of Russian websites, our approach can be used to estimate the influence of news websites more generally (\textit{e.g.}, {cnn.com, foxnews.com}). 

\vspace{2pt}
\noindent
\textbf{\textit{Responding to Propaganda.}}
As seen in this work, for websites like waronfakes.com, nearly 30\% of their content appears on Telegram at least a day before appearing on the website. By monitoring and understanding the Telegram ecosystem, researchers and fact-checkers can more quickly respond to topics that start on Telegram and make their way to other news sites. We argue that by ignoring Telegram, the research community has discounted a large and influential aspect of Russian propaganda pathways.  

\bibliography{references}
\end{document}